\documentclass[aps,prb,showpacs,superscriptaddress,twocolumn]{revtex4}
\usepackage[pdftex]{graphicx}  
\pdfoutput=1
\usepackage{dcolumn}   
\usepackage{bm}        
\usepackage{amssymb}   
\begin{document}

\title{Coexistence of two- and three-dimensional Shubnikov-de Haas oscillations in Ar$^+$-irradiated KTaO$_3$}
\author{S.~Harashima} 
\affiliation{Stanford Institute for Materials and Energy Sciences, SLAC National Accelerator Laboratory, Menlo Park, CA 94025, USA} 
\affiliation{Department of Applied Physics, The University of Tokyo, Bunkyo-ku, Tokyo 113-8656, Japan}
\author{C.~Bell} 
\affiliation{Stanford Institute for Materials and Energy Sciences, SLAC National Accelerator Laboratory, Menlo Park, CA 94025, USA}
\author{M.~Kim}
\affiliation{Stanford Institute for Materials and Energy Sciences, SLAC National Accelerator Laboratory, Menlo Park, CA 94025, USA}
\affiliation{Department of Advanced Materials Science, The University of Tokyo, Kashiwa, Chiba 277-8561, Japan}
\author{T.~Yajima}
\affiliation{Stanford Institute for Materials and Energy Sciences, SLAC National Accelerator Laboratory, Menlo Park, CA 94025, USA}
\affiliation{Department of Advanced Materials Science, The University of Tokyo, Kashiwa, Chiba 277-8561, Japan}
\author{Y.~Hikita} 
\affiliation{Stanford Institute for Materials and Energy Sciences, SLAC National Accelerator Laboratory, Menlo Park, CA 94025, USA}
\author{H.~Y.~Hwang} 
\affiliation{Stanford Institute for Materials and Energy Sciences, SLAC National Accelerator Laboratory, Menlo Park, CA 94025, USA}
\affiliation{Geballe Laboratory for Advanced Materials, Department of Applied Physics, Stanford University, Stanford, CA 94305, USA}
\date{\today}

\begin{abstract}
We report the electron doping in the surface vicinity of KTaO$_3$ by inducing oxygen-vacancies via Ar$^+$-irradiation.
The doped electrons have high mobility ($>$ 10$^4$ cm$^2$/Vs) at low temperatures, and exhibit Shubnikov-de Haas oscillations with both two- and three-dimensional components. A disparity of the extracted in-plane effective mass, compared to the bulk values, suggests mixing of the orbital characters. Our observations demonstrate that Ar$^+$-irradiation serves as a flexible tool to study low dimensional quantum transport in $5d$ semiconducting oxides.
\end{abstract}

\pacs{73.20.-r, 73.21.Fg, 68.47.Gh}
\maketitle
Transition metal oxides have been studied intensively due to their rich physics such as high temperature superconductivity and colossal magnetoresistance. Among various transition metal oxides, the high mobility $3d$ semiconductor SrTiO$_3$ (STO) has attracted much recent attention especially when it is confined in two dimensions.\cite{Ohtomo2004,UenoSTO2008,Kozuka2009,JalanStemmerPRB,Santander-Syro,Meevasana,PhysRevLett.106.136809} However, far fewer studies have investigated the related $5d$ compound KTaO$_3$ (KTO), even though it shares many fascinating properties of STO as well as distinct ones such as a lack of a structural phase transition at low temperatures.\cite{Samara1973} Stemming from the cubic crystal structure and the presence of the heavy element Ta, the conduction band minimum of KTO consists of a doubly degenerate light and heavy electron band with an additional spin-orbit split off band with a large spin orbit splitting $\sim$ 0.4 eV.\cite{Mattheiss1972} Recent efforts to understand the low dimensional properties of KTO have included subband structure studies at the cleaved KTO surface,\cite{King:2012} large Rashba spin orbit coupling found in field effect transistors (FETs),\cite{Nakamura:2009} and electric field induced superconductivity.\cite{Ueno:2011}

However, unlike thin films of doped STO which can be designed with atomic precision, the growth of chemically doped KTO thin films is more challenging due to the high volatility of K.\cite{Bae200451} As a result, studies of low dimensional transport in KTO are often limited to FET structures utilizing bulk single crystals.\cite{uenoAPL2004,Nakamura:2009,APEX.2.121103,Ueno:2011} More importantly, the maximum Hall mobility observed so far has not exceeded 7$\times$10$^{3}$ cm$^2$/Vs at temperatures $T<10$ K, and quantum oscillations have not yet been observed in these two-dimensional (2D) channels.
Here we report the realization of low dimensional conducting KTO by Ar$^+$-irradiation on the surface on undoped single crystals. The doped electrons have high mobilities (up to $2.4 \times 10^{4}$ cm$^2$/Vs) at $T = 2$ K.
The key result of the present study is that the longitudinal magnetoresistance clearly exhibits Shubnikov-de Haas quantum oscillations (SdHOs) which give vital information about the electronic structure of the electron gas. Surprisingly we find 2D and three-dimensional (3D) SdHO components in coexistence with one another, providing the opportunity to study low dimensional quantum transport in a $5d$ system.

\begin{figure}
	\centering
	\includegraphics[width=6cm]{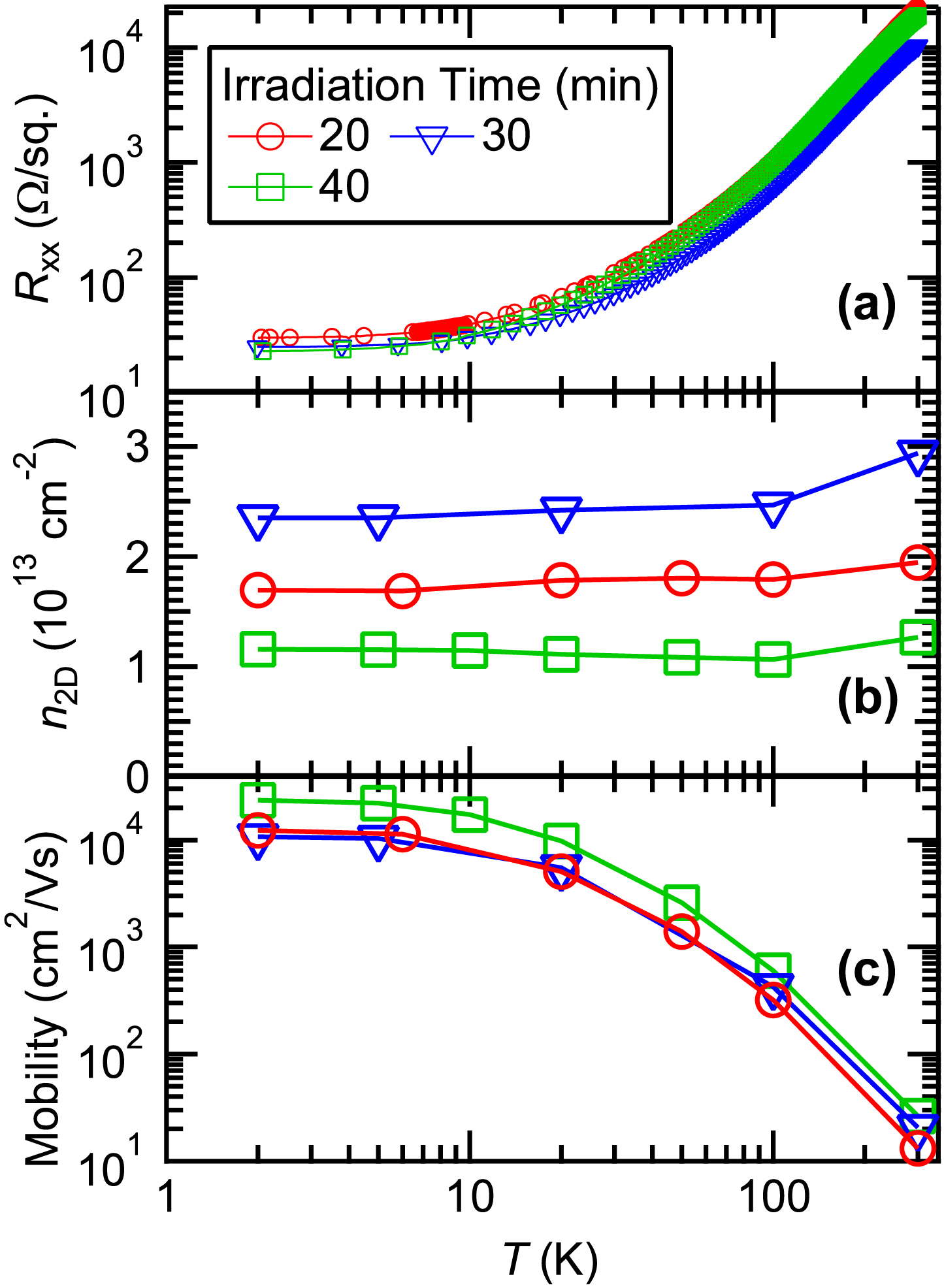}
	\caption{(Color online) The temperature dependence of (a) the sheet resistance $R_{xx}$, 
			(b) the sheet carrier density $n_{2D}$, and 
			(c) the Hall mobility $\mu$ of three samples with different Ar$^+$-irradiation times.}
	\label{Fig1}
\end{figure}%

Commercial $0.5$ mm thick single crystals (SurfaceNet GmbH) of undoped-KTO $\{$001$\}$ were cut into $3\times 3$ mm$^2$ pieces, and irradiated with Ar$^+$-ions at an acceleration voltage of 500 V at normal incidence. The etching vacuum chamber had a base pressure of 10$^{-5}$ Pa. The current of Ar$^+$-flux was 0.1 mA/cm$^2$. Samples were irradiated for the irradiation time $t$ in the range 1 min $\leq t\leq$ 40 min, on a water cooled sample holder. For relatively long $t$ ($>$ 10 min), the irradiation was done in intervals of 10 min, with a pause of 5 min, in order to avoid significant heating of the substrates. After the Ar$^+$-irradiation, the sheet resistance and the Hall effect were measured in a pumped He-4 cryostat using a standard four probe method with wire-bonded Al contacts. An $in$-$situ$ horizontal rotator with an accuracy better than $0.1^{\circ}$ was used to vary the inclination angle $\theta$ between the sample normal and the magnetic field in the cryostat.

Above a critical irradiation time of $t = 20$ min, all samples became metallic. For 20 min $\le t \le$ 40 min the sheet carrier density $n_{2D}$ was of the order of 10$^{13}$ cm$^{-2}$. Fig.~\ref{Fig1} summarizes the temperature dependence of the longitudinal sheet resistance $R_{xx}$, $n_{2D}$ and the Hall mobility $\mu$ for three representative samples. $R_{xx}$ monotonically decreases with decreasing $T$ down to 2 K and $n_{2D}(T)$ is nearly flat without strong freeze out of electrons, which is a
typical feature of degenerate semiconductors.
$n_{2D}$ does not monotonically increase with increasing $t$, as also observed in the case of Ar$^+$-irradiated STO.\cite{SatoJJAP2007,Reagor2005,GrossJAP2011} In that case the saturation of the sheet carrier density was attributed to the so-called ``high dose limit",\cite{Nastasi} where the preferential etching of oxygen atoms and the physical removal of other atoms (Sr and Ti) are in dynamic balance, leading to a sheet density of oxygen vacancies independent of $t$. Notably, in the case of STO, $n_{2D}$ in the high dose limit was significantly larger ($\sim$ 10$^{14}$ cm$^{-2}$) for the same irradiation conditions.\cite{SatoJJAP2007} The relatively low $n_{2D}$ in KTO implies that the the physical removal of K and Ta is stronger compared to Sr and Ti, presumably due to a combination of the high volatility of K and the
large sputter yield of Ta,\cite{SigmundSputteringTheory} therefore the preferential etching of O atoms is weaker compared to the case in Ar$^+$-irradiated STO.

Remarkably, $\mu$ reaches 2.4$\times$10$^{4}$ cm$^2$/Vs at $T = 2$ K for the $t = 40$ min sample. This value is significantly higher than any reported values of Ar$^+$-irradiated STO,\cite{Reagor2005,SatoJJAP2007,Schultz:2007,herranz:103704,Guerrero:2010,Ngai:2010,Ngai:2011,Guerrero2011,Bruno:2011}
 or in KTO FETs.\cite{uenoAPL2004,Nakamura:2009,Ueno:2011} Moreover, this value is comparable to the highest $\mu$ found in bulk KTO (2.3$\times$10$^4$ cm$^2$/Vs).\cite{WempleTransport} This can be understood if the dopant oxygen vacancies distributes into the substrate and hence the conduction electrons in the quantum well are free from significant surface or interface scattering. These data suggest that Ar$^+$-irradiation is a flexible way to induce carriers in KTO.

\begin{figure}
	\centering
	\includegraphics[width=6cm]{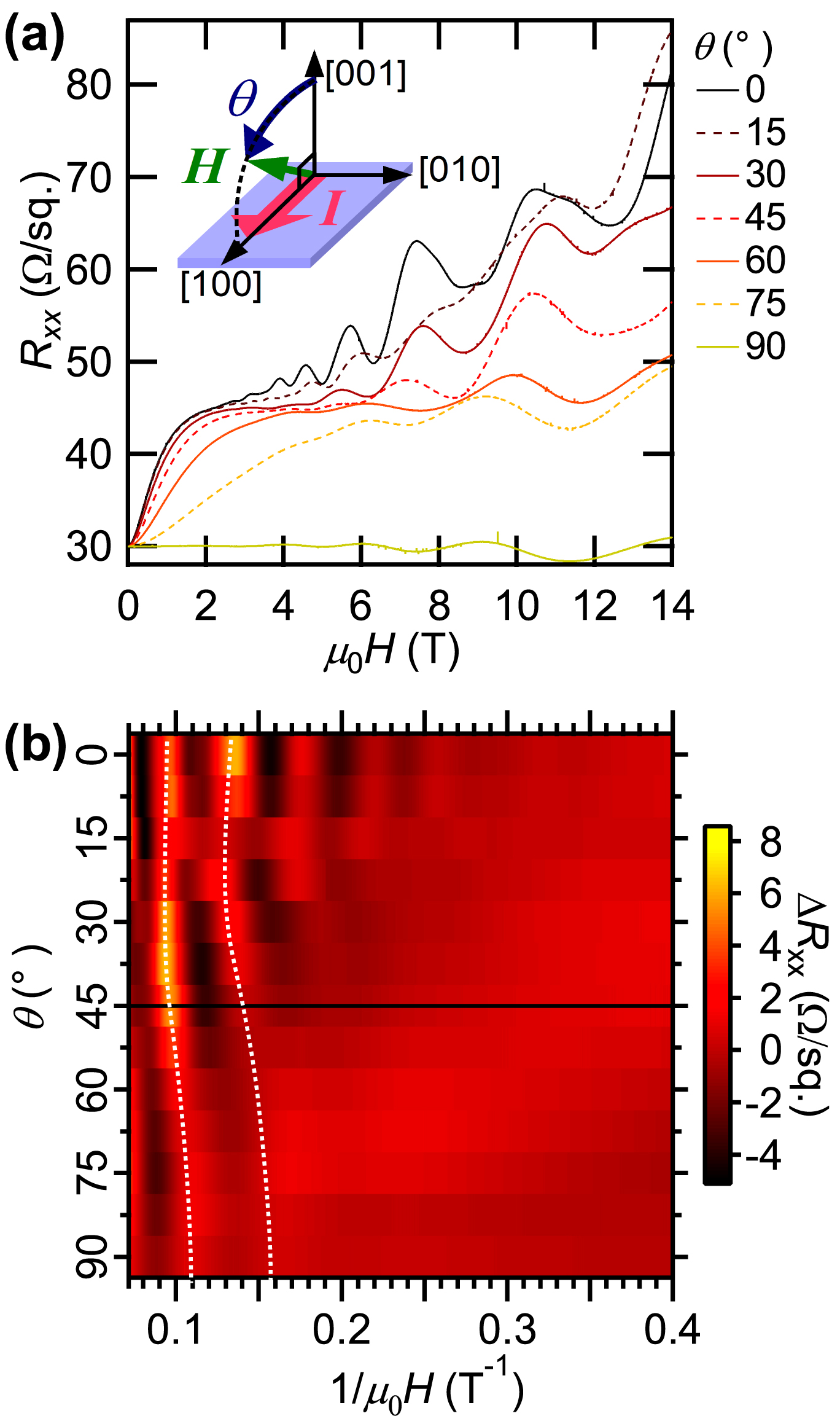}
	\caption{(Color online)(a) The longitudinal magnetoresistance $R_{xx}$ at $T=2$ K for various inclination angles $\theta$ of the magnetic field. 
			The inset diagram shows the geometry of the current $I$, the magnetic field $H$, and the inclination angle $\theta$ with the crystallographic directions.
			(b) Image plot of the oscillatory component of the magnetoresistance $\Delta R_{xx}$, 
			extracted by subtracting a second-order polynomial fitting to the $R_{xx}(H)$ over 
			a magnetic field range $1.5$ T $<\mu_0H<14$ T. The dotted lines are a guide to the eye showing the angular dependence of two peak positions.
			}
	\label{Fig2}
\end{figure}%

In general, in such high mobility electron gases, the long mean free time enables the observation of magnetic quantum oscillations.\cite{SchoenbergBook}
Indeed, we observed SdHOs in all metallic samples for $T< 5$ K. Hereafter, we focus on the $t = 20$ min sample for clarity. 
The magnetoresistance as a function of $\theta$ at $T=2$ K is shown in Fig.~\ref{Fig2}(a) for several representative $\theta$. As shown in the inset of Fig.~\ref{Fig2}(a), 
the inclination angle $\theta$ is defined such that the direction of the magnetic field $H$ is parallel to the sample normal for $\theta = 0^{\circ}$, and parallel to the current for $\theta= 90^{\circ}$, which is itself along a $<$100$>$ direction in-plane. Clear oscillations of $R_{xx}$ are observed superimposed on a positive magnetoresistance background. This background, which we ascribe to the classical orbital effect,\cite{Ngai:2010} was subtracted for each data set using a second-order polynomial fitting over the range $1.5$ T $<\mu_0H<14$ T.

As shown in Fig.~\ref{Fig2}(b), the extracted oscillating part of $R_{xx}$, $\Delta R_{xx}$ exhibits SdHOs for all $\theta$, meaning that there is a 3D Fermi surface in this electron gas. However, the symmetry of these 3D SdHOs is not what is expected for bulk KTO, whose crystal structure, and hence Fermi surface, has a cubic symmetry.\cite{Mattheiss1972, Uwe1979} In the SdHO data, this bulk cubic symmetry of the Fermi surface should be reflected as a line symmetry about $\theta = 45^\circ$, which is clearly not observed in Fig.~\ref{Fig2}(b), where the angular dependences of two peak positions at relatively high fields are highlighted by dotted lines. 

\begin{figure}
	\centering
	\includegraphics[width=6cm]{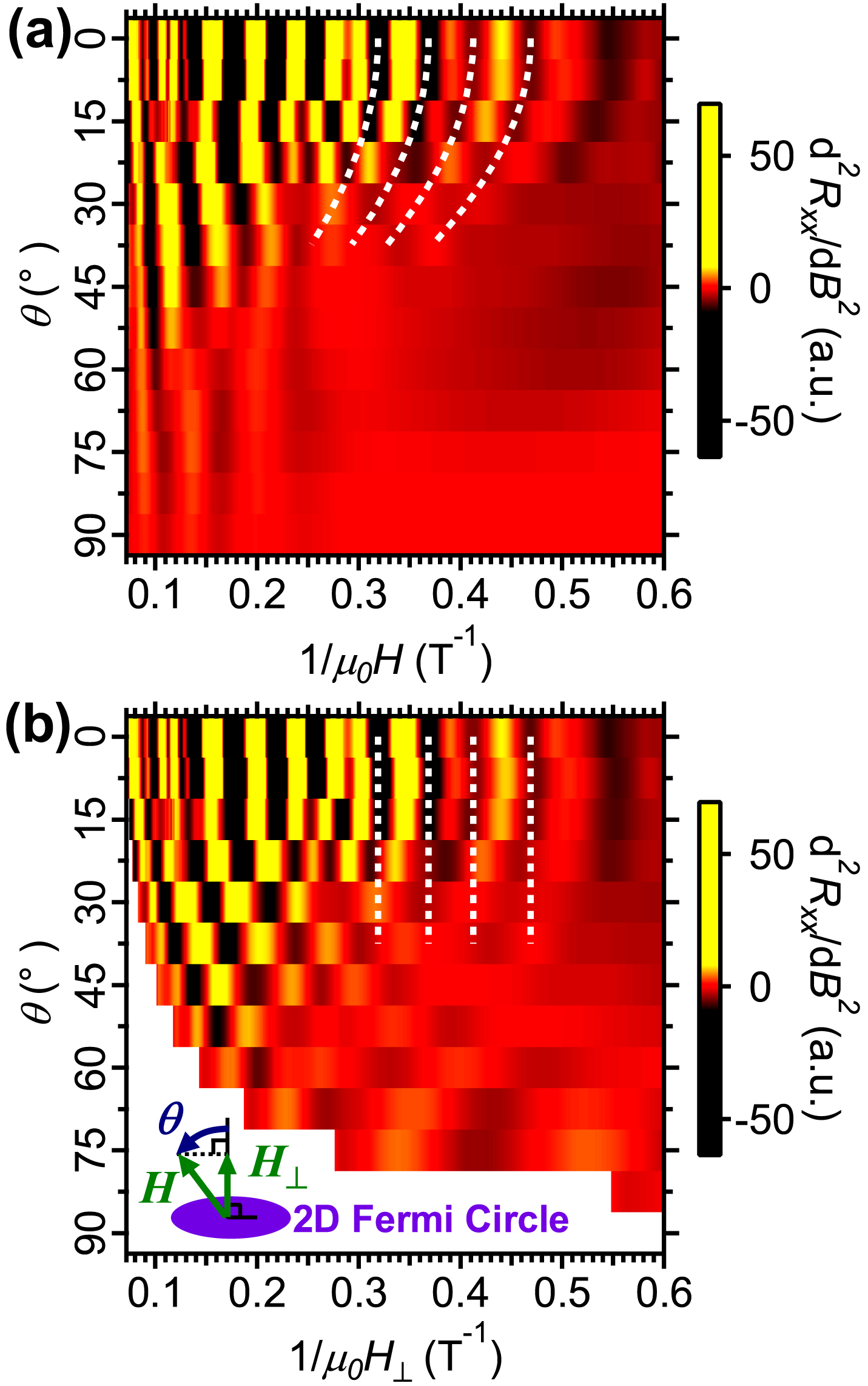}
	\caption{(Color online)(a) Image plot of the second derivative of the magnetoresistance $\mathrm{d}^2R_{xx}/\mathrm{d}B^2$ for various $\theta$. 
			(b) $\mathrm{d}^2R_{xx}/\mathrm{d}B^2$ scaled by the perpendicular component of the magnetic field $\mu _0H_\perp$.
			The dashed lines show the scaling of $H_\perp(\theta)$.
			The white area in (b) is not measured in this experiment.
			The inset is a schematic diagram of $\theta$, $H$, $H_{\perp}$ and a 2D Fermi circle.}
	\label{Fig3}
\end{figure}

This deviation from bulk KTO symmetry is more evident in the second derivative $\mathrm{d}^2R_{xx}/\mathrm{d}B^2$, as shown in Fig.~\ref{Fig3}(a). The same data, scaled by the perpendicular magnetic field component $H_\perp=H\mathrm{cos}(\theta)$ are shown in Fig.~\ref{Fig3}(b). In Figs.~\ref{Fig3}(a,b), the dashed lines show the scale of $H_\perp(\theta)$. Here, the use of the derivative amplifies the SdHOs at relatively low $H$, revealing additional components which were difficult to resolve in the $\Delta R_{xx}$ data, especially in the regime $\theta<30^\circ$ and $1/\mu_0 H>0.3$ T$^{-1}$. In addition to the clear lack of symmetry about $\theta=45^\circ$, already noted above, strikingly we can also observe SdHO peaks which scale with $H_\perp(\theta)$, and follow the dashed lines as shown in Fig.~\ref{Fig3}(b), giving clear evidence for 2D electronic states. This strongly suggests the coexistence of 2D and 3D Fermi surfaces in the vicinity of the irradiated KTO surface.

\begin{figure}
	\centering
	\includegraphics[width=8cm]{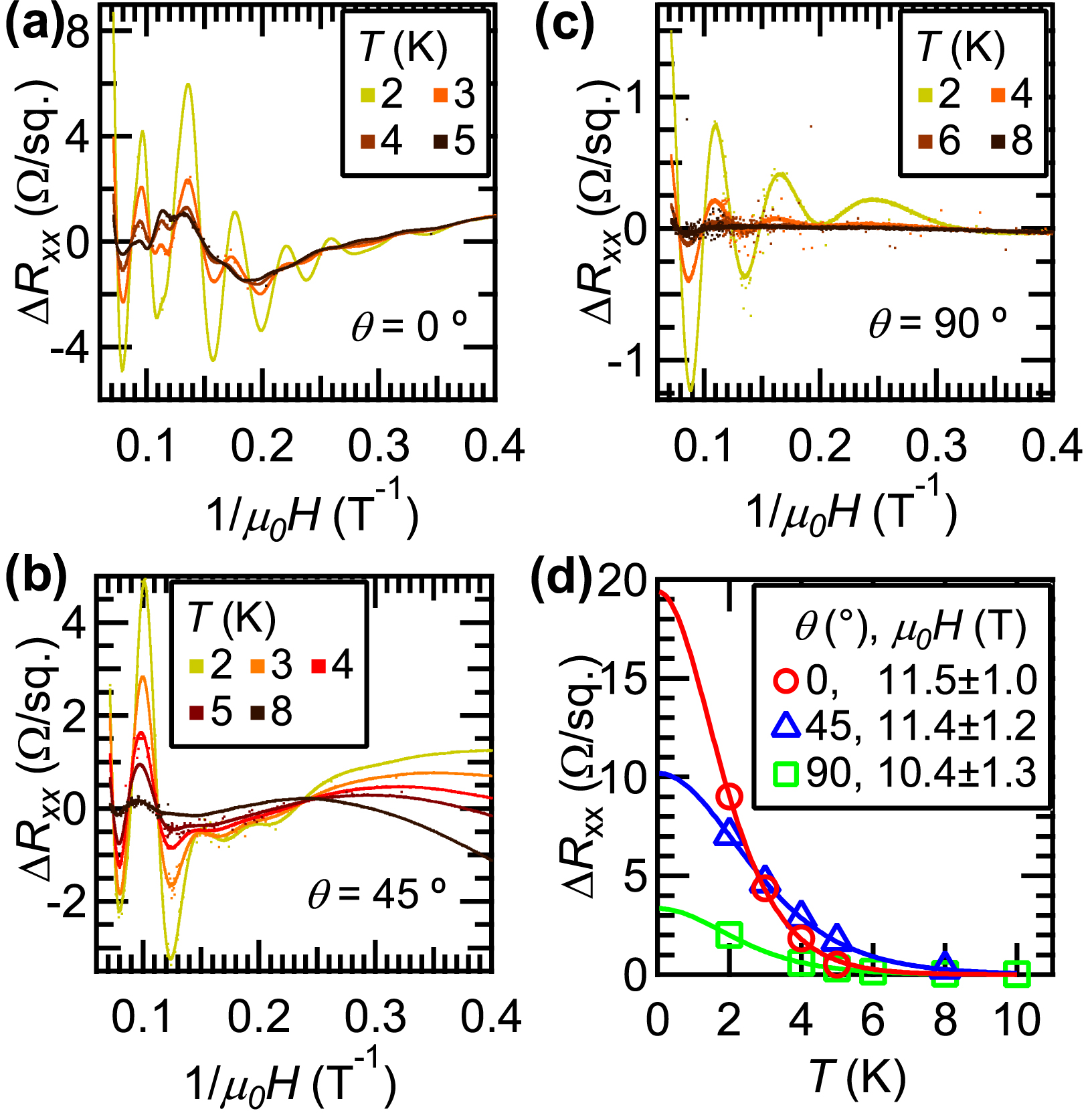}
	\caption{(Color online) $\Delta R_{xx}$ at various temperatures for $\theta=$ 0$^\circ$ (a), 45$^\circ$ (b) and 90$^\circ$ (c).
			The background magnetoresistance was subtracted by a fitting second order polynomial to $R_{xx}(H)$.
			The temperature dependence at specific magnetic fields are plotted in (d).
			Solid lines are the best fits to Eq.~(\ref{LKTemp}).
			Fitting results of $m^*$ are (0.90$\pm$0.08)$m_0$, (0.60$\pm$0.06)$m_0$, and (0.65$\pm$0.08)$m_0$ 
			for $\theta=0^\circ$, 45$^\circ$ and 90$^\circ$, respectively.
			}
	\label{Fig4}
\end{figure}

Next we investigated the temperature dependence of the SdHOs, in order to extract the effective mass of the carriers.
The SdHOs at various temperatures were measured  at $\theta=$ 0$^\circ$, 45$^\circ$ and 90$^\circ$, as shown in Fig.~\ref{Fig4}(a-c).
Here, the background magnetoresistance was subtracted at each temperature and angle in the same manner as Fig.~\ref{Fig2}(b).
The SdHO amplitude was extracted by focusing on one half-period of the 
oscillations, and determining the peak-to-peak value. The magnetic field 
ranges of these half-periods were 11.5$\pm$1.0 T, 11.4$\pm$1.2 T and 10.4$\pm$1.3 T for $\theta=0^\circ$, 45$^\circ$ and 90$^\circ$, respectively. The temperature dependences of these amplitudes were fitted using the standard Lifshitz-Kosevich formula, 
\begin{equation}
\label{LKTemp}
R_T = A\frac{2 \pi ^2 k_B T m^* / e \hbar B}{\mathrm{sinh} \left( 2 \pi ^2 k_B T m^* / e \hbar B\right)},
\end{equation}
where $A$ is the proportional constant, $k_B$ is the Boltzmann constant, $m^*$ is the effective mass of the electron,
$\hbar$ is the reduced Planck constant, and $B$ is the magnetic flux density.\cite{Sergio2001} The amplitude data and corresponding fits are shown in Fig.~\ref{Fig4}(d). These fits resulted in $m^*$ values of (0.90$\pm$0.08)$m_0$, (0.60$\pm$0.06)$m_0$, and (0.65$\pm$0.08)$m_0$ 
for $\theta=0^\circ$, 45$^\circ$ and 90$^\circ$, respectively, where $m_0$ is the bare electron mass. While $m^*$ at $\theta=45^\circ$ and 90$^\circ$ are in good agreement with a previously reported value of the light electron mass of bulk KTO (0.55$\pm$0.05)$m_0$,\cite{Uwe1979} we find that $m^*$ at $\theta=$ 0$^\circ$ is significantly larger, and comparable to the bulk heavy mass (0.80$\pm$0.05)$m_0$.\cite{Uwe1979}
 
In order to understand all of the above results, it is necessary to consider 
both the structural and electronic contributions to the formation of the 
electron gas. Here, one may consider structural changes near the surface 
region and the resulting change of the Fermi surface shape, associated with 
the Ar$^+$-irradiation. Indeed, Kan {\it et al.}~\cite{Kan2007} reported that 
the lattice constant of Ar$^+$-irradiated STO expands in depth direction due 
to the presence of oxygen vacancies. Such a tetragonal distortion of the lattice 
would result in $less$ overlap of the neighboring Ta 5$d$-$t_{2g}$ orbits in the depth direction, distorting the cubic 
symmetric Fermi surface into an ellipsoid with its major axis along the 
out-of-plane direction. However, in this case the Fermi surface extremal 
cross-sectional area is small (large) when $\theta = 0^{\circ}$ ($90^{\circ}$
), in disagreement with the data shown in Fig.~\ref{Fig2}(b), where the peak 
positions denoted by the dotted lines move to $larger$ $(\mu_0H)^{-1}$ with 
increasing $\theta$. In addition to the angular dependence of the peak 
positions, the tetragonal distortion also cannot explain the difference of 
$m^*$ for $\theta=0^\circ$ and $90^\circ$.
$m^*$ expected from the ellipsoidal Fermi surface should be light (heavy) for $\theta=0^\circ$ ($90^\circ$), which disagrees with the fitting results to the temperature dependences of the SdHOs.
Hence, in order to understand the observed SdHOs, we must instead consider the quantum well structure itself.
 
Due to the strongly depth dependent density of induced oxygen vacancies,\cite{GrossJAP2011} the sub-band splitting in the quantum well is expected to be rather non-uniform as a function of sub-band index. If the quantization energy is larger than the disorder broadening in some sub-bands, 2D electron gas behavior is observed. On the other hand,
other sub-bands, with smaller quantization energies may be disorder broadened and form a 3D Fermi surface. Such a coexistence of different dimensionalities has been observed in a wide parabolic GaAs-based system, as discussed by Sergio {\it et al.}\cite{Sergio2001} This sub-band structure naturally explains the two types of the observed SdHOs. The temperature dependence data show that the 2D electrons, which are probed at $\theta=0^{\circ}$, are relatively heavy, which is consistent with a complex mixture of different bulk orbital characteristics for the various sub-bands inside the well.\cite{Popovic2008, King:2012} 

We note that only a small fraction of total carriers measured by the Hall effect showed 2D SdHOs. These 2D oscillations, denoted by the dashed lines in Fig.~\ref{Fig3}, have an oscillation frequency of $\sim 20.2$ T. Assuming spin degeneracy of the Landau levels, this frequency corresponds to a sheet carrier density 9.8$\times$10$^{11}$ cm$^{-2}$, representing only 5.8 \% of the total Hall carrier density of 1.7$\times10^{13}$ cm$^{-2}$. The remaining conduction electrons show either 3D SdHOs or no oscillatory behavior. Similar behavior has also be observed in a range of STO-based heterostructures.\cite{Kozuka2009,KimPRL2011,BenShalomSdH,SonStemmerNMat,JalanStemmerPRB,CavigliaSdH}

Our observation of coexisting electron dimensionalities implies the possibility of similar physics in Ar$^+$-irradiated STO. Indeed, the length scale of the oxygen vacancy distribution in the depth direction of STO has been reported in several works with a wide variation. For example, a synchrotron X-ray diffraction study showed a 21 nm thick lattice distortion due to oxygen vacancies,\cite{Kan2007} and photoluminescence dynamics suggested a 61 nm deep electron gas,\cite{YasudaPLDynamics2008} whereas conducting tip atomic force microscopy and positron annihilation spectroscopy studies showed a much larger length scale $\sim$ 1 $\mu$m.\cite{herranz:103704} These measurements may not be inconsistent, if the different characterization methods are probing the different aspects
of a multi-component electron gas. A drawback in the case of STO is that the relatively small $\mu$ hinders the direct clarification of the electronic structure using SdHOs, in contrast to the current experiments.

In summary, we successfully doped high mobility electrons by inducing oxygen vacancies in the surface vicinity of KTO by means of Ar$^+$-irradiation, enabling SdHOs with multiple components to be observed. The angular- and temperature-dependence of the SdHOs showed clear deviations from the bulk character, demonstrating the coexistence of 2D and 3D electron states. These results emphasize the power of Ar$^+$-irradiation as a flexible tool for studying low temperature quantum transport in $d$-electron systems.

The authors thank M.~Lippmaa for experimental assistance 
and acknowledge support by the Department of Energy, Office of Basic Energy Sciences, Materials Sciences and Engineering Division, under contract DE-AC02-76SF00515.

\end{document}